\documentclass[12pt,preprint]{aastex}  
\usepackage{graphicx}
\usepackage{txfonts}
\newcommand{\va}{v_{\mathrm{A}}}
\newcommand{\vai}{v_{\mathrm{Ai}}}
\newcommand{\vae}{v_{\mathrm{Ae}}}

\newcommand{\pd}{\partial}

\begin{document}

\title{KELVIN-HELMHOLTZ INSTABILITY IN CORONAL MAGNETIC FLUX TUBES DUE TO AZIMUTHAL SHEAR FLOWS}

\shorttitle{K.H INSTABILITY IN CORONAL LOOPS}

   \author{R. Soler$^1$, J. Terradas$^{1,2}$, R. Oliver$^1$, J. L. Ballester$^1$, and M. Goossens$^2$}
   \affil{$^1$ Departament de F\'isica, Universitat de les Illes Balears,
              E-07122, Palma de Mallorca, Spain}
              \email{roberto.soler@uib.es}

  \affil{$^2$ Centre for Plasma Astrophysics and Leuven Mathematical Modeling and Computational Science Centre, K. U. Leuven, Celestijnenlaan 200B, 3001 Heverlee, Belgium}

  \begin{abstract}
Transverse oscillations of coronal loops are often observed and have been theoretically interpreted as kink magnetohydrodynamic (MHD) modes. Numerical simulations by Terradas et al. (2008, ApJ 687, L115) suggest that shear flows generated at the loop boundary during kink oscillations could give rise to a Kelvin-Helmholtz instability (KHI). Here, we investigate the linear stage of the KHI in a cylindrical magnetic flux tube in the presence of azimuthal shear motions. We consider the basic, linearized MHD equations in the $\beta = 0$ approximation, and apply them to a straight and homogeneous cylindrical flux tube model embedded in a coronal environment. Azimuthal shear flows with a sharp jump of the velocity at the cylinder boundary are included in the model. We obtain an analytical expression for the dispersion relation of the unstable MHD modes supported by the configuration, and compute analytical approximations of the critical velocity shear and the KHI growth rate in the thin tube limit. A parametric study of the KHI growth rates is performed by numerically solving the full dispersion relation. We find that fluting-like modes can develop a KHI in time-scales comparable to the period of kink oscillations of the flux tube. The KHI growth rates increase with the value of the azimuthal wavenumber and decrease with the longitudinal wavenumber. However, the presence of a small azimuthal component of the magnetic field can suppress the KHI. Azimuthal motions related to kink oscillations of untwisted coronal loops may trigger a KHI, but this phenomenon has not been observed to date. We propose that the azimuthal component of the magnetic field is responsible for suppressing the KHI in a stable coronal loop. The required twist is small enough to prevent the development of the pinch instability.

  \end{abstract}



\section{INTRODUCTION}

The phenomenon of transverse coronal loop oscillations has received much attention during the last decade since the first observational reports \citep[e.g.,][]{asch,naka1} and their subsequent theoretical interpretation in terms of kink magnetohydrodynamic (MHD) modes \citep[e.g.,][]{naka2,rudermanroberts,goossens02}. The reader is referred to recent reviews, and references therein, regarding the modeling of transverse loop oscillations in terms of MHD eigenmodes \citep{rudermanrobertus}, their time-dependent excitation \citep{terradasreview}, and their non-linear evolution \citep{ofman09}. 

\citet{terradas} studied 3D numerical simulations of non-linear kink oscillations of a straight, untwisted coronal magnetic flux tube. These authors found that, after kink motions are excited, small length-scale disturbances appear at the tube boundary and grow rapidly in time. The inclusion of a transversely inhomogeneous transitional layer between the flux tube and the external medium does not suppress the apparition of these small-scale disturbances, but they develop more slowly than in the sharp transition situation. Although \citet{terradas} qualitative interpreted this phenomenon in terms of a Kelvin-Helmholtz instability (KHI), which would be triggered by shear flows at the tube boundary, a more detailed investigation of the instability regimes and growth rates is needed. This can be performed in two different ways. The first option is to study the non-linear evolution of the KHI by numerically solving the full 3D non-linear MHD equations as in \citet{terradas}, but with a very large spatial resolution able to correctly describe the small spatial-scales generated at the tube boundary. This procedure requires an enormous computational effort. The second option is to attack the problem analytically by restricting ourselves to the initial stage of the KHI and considering its linear evolution in a simplified model that includes the basic ingredients of the actual situation. By means of an analytical work, a more in-depth investigation of the effects and parameters involved in the generation of the KHI is possible. We adopt this second approach in the present work. 

To do so, we have to identify the basic ingredients that the model must contain. We consider a straight cylindrical flux tube embedded in a coronal environment to represent a coronal loop. It is well-known that for kink oscillations of the flux tube, the azimuthal component of the velocity perturbation has a finite jump at the cylinder boundary \citep[see, e.g., Figure~1(b) of][]{goossens}. We claim that this jump of the azimuthal velocity is responsible for the observed KHI in the simulations by \citet{terradas}. So, we take the presence of azimuthal shear flows at the cylinder into account to represent those azimuthal motions generated by the kink oscillations themselves. However, to consider the actual azimuthal velocity profile of the kink mode perturbation (expressed in terms of Bessel functions) makes the problem not analytically tractable. Instead, we assume here a simpler profile for the azimuthal velocity which allows us to proceed analytically, but keeping the presence of a sharp velocity discontinuity at the cylinder boundary as the key ingredient of the real situation. We also neglect the dependence of the flow on the longitudinal direction. In addition, we assume that the azimuthal shear flows are not time-dependent. This condition restricts ourselves to KHI growth rates much smaller than the kink mode period, but it allows us to perform an analysis in terms of normal modes.

The effect of mass flow on coronal loop oscillations has been studied by a number of authors, but they mainly focused on longitudinal flows \citep[e.g.,][among other works]{goossens92, robertusflow, terra, terradasflow}. Applications to filament thread oscillations have also been performed \citep{solerflow,solermulti}. Concerning the stability of these flows, both the KHI and the resonant instability have been investigated \citep[e.g.,][]{rae,andries1,andries2,taroyan,holzwarth}. Regarding azimuthal shear flows, there are several previous investigations in the context of astrophysical and laboratory plasmas that are relevant for the present application.  A very comprehensive investigation of a similar configuration to that adopted here but in the context of tokamaks can be found in \citet{bondeson}. \citet{bodo1, bodo2} studied the stability of cylindrical jets in the presence of both azimuthal and longitudinal flows. While they first restricted themselves to the axisymmetric case in \citet{bodo1}, they later extended their investigation to non-axisymmetric perturbations in \citet{bodo2}. These authors obtained an analytical expression for the dispersion relation of the unstable modes and computed the instability growth rates under astrophysical jet conditions. \citet{ogilvie} studied the stability of a differentially rotating accretion disk in the presence of an azimuthal magnetic field, and \citet{kolesnikov} investigated the KHI due to helical flow around a photospheric magnetic flux tube. To our knowledge, no similar analytical work applied to the context of kink oscillations of coronal loops is found in the literature.

\section{BASIC EQUATIONS}
\label{sec:math}

Our configuration is composed of a homogeneous, straight, and cylindrical magnetic flux tube of radius $a$ embedded in a homogeneous coronal environment. The geometry of the model allows us to use cylindrical coordinates, namely $r$, $\varphi$, and $z$, for the radial, azimuthal, and longitudinal coordinates, respectively. In the following expressions, a subscript 0 indicates equilibrium quantities while we use subscripts $\rm i$ and $\rm e$ to explicitly denote internal and external quantities, respectively. The magnetic field is taken homogeneous and orientated along the cylinder axis, $\mathbf{B}_0 = B_0 \hat{e}_z$, with $B_0$ constant everywhere. We assume a step profile for the density, with $\rho_{\rm i}$ and $\rho_{\rm e}$ the internal and external densities, respectively. We consider an azimuthal shear flow in the equilibrium configuration, $ \mathbf{U}_0 = U_0 \hat{e}_\varphi$, with $U_0$ given by
\begin{equation}
 U_0 = \left\{ \begin{array}{lcl}
                         v_0 r, & \textrm{if} & r \leq a, \\
			0, & \textrm{if} & r > a.
                        \end{array}
  \right.
\end{equation}
Thus, $v_0 a$ is the velocity jump at the cylinder boundary. We are aware that the selected flow profile is substantially different from the kink mode azimuthal velocity. However, our flow profile keeps the key ingredient of the actual kink mode profile, i.e., the presence of a sharp jump at the cylinder edge. The KHI is mainly governed by this jump of the azimuthal velocity, while the precise flow profile within the flux tube would only have a minor influence on the growth rates. In addition, the present profile allows us to proceed analytically. Here, and as in \citet{bodo1,bodo2}, we assume that $v_0 a \ll \vai$, with $\vai$ the internal Alfv\'en speed (defined below), which allows us to neglect the effect of the centrifugal force over the equilibrium total pressure gradient and therefore the equilibrium quantities can be taken constant in the radial direction. Since part of the following treatment is equivalent to that of \citet{bodo1,bodo2}, we will refer to their works when appropriate. 

We consider the linearized MHD equations for an ideal plasma in the $\beta = 0$ approximation, namely
\begin{equation}
\rho_0 \left( \frac{\pd \mathbf{v}_1}{\pd t} + \mathbf{U}_0 \cdot \nabla \mathbf{v}_1 + \mathbf{v}_1 \cdot \nabla \mathbf{U}_0 \right) = \frac{1}{\mu} \left( \nabla \times \mathbf{B}_1 \right) \times  \mathbf{B}_0, \label{eq:momentum}
\end{equation}
\begin{equation}
 \frac{\pd \mathbf{B}_1}{\pd t} = \nabla \times \left( \mathbf{U}_0 \times \mathbf{B}_1 \right)  + \nabla \times \left( \mathbf{v}_1 \times \mathbf{B}_0 \right),\label{eq:induction}
\end{equation}
along with the condition $\nabla \cdot {\mathit {\bf B}}_1 = 0$. In these equations ${\mathit {\bf B}}_1 = \left( B_r , B_\varphi, B_z \right)$, ${\mathit {\bf v}}_1 = \left( v_r , v_\varphi, v_z \right)$ are the components of the magnetic field and velocity perturbations, respectively, while  $\mu$ is the vacuum magnetic permeability. Next, we write the perturbations proportional to $\exp \left( i m \varphi + i k_z z - i \omega t \right)$, where $\omega$ is the oscillatory frequency, and $m$ and $k_z$ are the azimuthal and longitudinal wavenumbers, respectively. The effect of line-tying of the flux tube footpoints in the photosphere is introduced by a quantization of the longitudinal wavenumber,
\begin{equation}
 k_z = \frac{n \pi}{L}, \quad \textrm{for}\quad n =1,2,3\dots, \label{eq:kz}
\end{equation}
with $L$ the cylinder length. In this model, $n=1$ corresponds to the fundamental mode. Then, Equations~(\ref{eq:momentum})--(\ref{eq:induction}) become,
\begin{equation}
  \Omega v_r =  2 i v_0 v_\varphi + \frac{\va^2}{B_0} \left( k_z B_r + i B_z' \right), \label{eq:vr}
\end{equation}
\begin{equation}
  \Omega v_\varphi = - 2 i v_0 v_r + \frac{\va^2}{B_0} \left( k_z B_\varphi -  \frac{m}{r} B_z \right),\label{eq:vf}
\end{equation}
\begin{equation}
 \Omega B_r = - B_0 k_z v_r,\label{eq:br}
\end{equation}
\begin{equation}
 \Omega B_\varphi = - B_0 k_z v_\varphi,\label{eq:bf}
\end{equation}
\begin{equation}
 \Omega B_z = - i B_0 \left( v_r' + \frac{1}{r} v_r + \frac{i m}{r} v_\varphi \right).\label{eq:bz}
\end{equation}
Since we assume the $\beta = 0$ approximation, we also have $v_z = 0$. In Equations~(\ref{eq:vr})--(\ref{eq:bz}), the prime denotes a radial derivative, $\va^2 = B_0^2/\sqrt{\mu \rho_0}$ is the Alfv\'en speed squared, and $\Omega = \omega - m U_0 / r$ is the so-called Doppler-shifted frequency. Note that $\Omega_{\rm i} = \omega - m v_0$ within the flux tube and $\Omega_{\rm e} = \omega$ in the coronal medium, meaning that in both regions the Doppler-shifted frequency is independent of $r$ for the given flow profile. Equations~(\ref{eq:vr})--(\ref{eq:bz}) can be combined into a single equation for the total pressure perturbation, $P_{\rm T} = B_0 B_z / \mu$, namely
\begin{equation}
 P_{\rm T}'' + \frac{1}{r} P_{\rm T}' + \left( k^2 - \frac{m^2}{r^2} \right)  P_{\rm T} = 0, \label{eq:basic}
\end{equation}
with
\begin{equation}
 k^2 = \frac{\Omega^2 - k_z^2 \va^2}{\va^2} - \frac{4 v_0^2 \Omega^2}{\left( \Omega^2 - k_z^2 \va^2 \right) \va^2}. \label{eq:kk}
\end{equation}
Note that Equation~(\ref{eq:basic}) is equivalent to Equation~(2.3) of \citet{bodo1} if the effect of the longitudinal flow is dropped from their expression. The Lagrangian radial displacement, $\xi_r = i v_r / \Omega$, is related to the total pressure perturbation as follows,
\begin{equation}
 \xi_r = \frac{\left[\left( \Omega^2 - k_z^2 \va^2 \right)  P_{\rm T}' - 2 v_0 \Omega \frac{m}{r} P_{\rm T} \right]}{\rho_0 \left[ \left( \Omega^2 - k_z^2 \va^2 \right)^2 - 4 v_0^2 \Omega^2  \right]}.
\end{equation}

The general solution of Equation~(\ref{eq:basic}) in the internal medium is
\begin{equation}
 P_{{\rm Ti}} =A_1 J_{m} \left( k_{\rm i} r \right) + A_2 Y_{m} \left( k_{\rm i} r \right),
\end{equation}
where $J_{m} $ and $Y_{m} $ are the Bessel functions of the first and second kind of order $m$, respectively \citep{abramowitz}, while $A_1$ and $A_2$ are constants, and $k_{\rm i}$ is given by Equation~(\ref{eq:kk}). We seek for regular solutions at $r=0$, so we impose $A_2 = 0$. On the other hand, the general expression for the total pressure perturbation in the corona is
\begin{equation}
 P_{{\rm Te}} = A_3 H_m^{\left( 1 \right)} \left( k_{\rm e} r \right) + A_4 H_m^{\left( 2 \right)} \left( k_{\rm e} r \right) ,
\end{equation}
with $H_m^{\left( 1 \right)}$ and $H_m^{\left( 2 \right)}$ the Hankel functions of the first and second kind of order $m$, respectively. In addition, $A_3$ and $A_4$ are constants, and $k_{\rm e}$ is defined as
\begin{equation}
 k_{\rm e}^2 = \frac{\omega^2 - k_z^2 \vae^2}{\vae^2}. \label{eq:ke}
\end{equation}
The condition for outgoing waves is fulfilled by setting $A_4 = 0$ and selecting the appropriate branch of $k_{\rm e}$ such that $\Re \left( k_{\rm e} / \omega \right) > 0$ \citep[see details in, e.g.,][]{wilson,cally}. 

The dispersion relation is obtained by imposing the continuity of both $P_{\rm T}$ and $\xi_r$ at the cylinder boundary, i.e., $r=a$,
\begin{equation}
  \frac{k_{\rm e}}{\rho_{\rm e} \left( \omega^2 - k_z^2 \vae^2 \right)}\frac{H_m^{' \left( 1 \right)} \left( k_{\rm e} a \right)}{H_m^{\left( 1 \right)} \left( k_{\rm e} a \right)}  = \frac{\left( \Omega_{\rm i}^2 - k_z^2 \vai^2 \right)  k_{\rm i} \frac{J_m' \left( k_{\rm i} a \right)}{J_m \left( k_{\rm i} a \right)} - 2 v_0 \Omega_{\rm i} \frac{m}{a} }{\rho_{\rm i} \left[ \left( \Omega_{\rm i}^2 - k_z^2 \vai^2 \right)^2 - 4 v_0^2 \Omega_{\rm i}^2  \right]}. \label{eq:reldisper}
\end{equation}
Equation~(\ref{eq:reldisper}) is the basic dispersion relation whose solutions we will discuss in this investigation, and is equivalent to Equation~(2.10) of \cite{bodo1} although with a different notation. Note that in the absence of flow, Equation~(\ref{eq:reldisper})  reduces to the well-known dispersion relation of MHD waves in a magnetic cylinder \citep[e.g.,][]{edwinroberts,goossens}. In order to characterize the solutions of Equation~(\ref{eq:reldisper}), it is useful to consider their wave properties. In the general situation and for fixed and real $k_z$, $m$, and $v_0$, complex values of the frequency, $\omega = \omega_{\rm R} + i \gamma$, are expected. If $\gamma < 0$ we have a damped mode, whereas $\gamma > 0$ corresponds to a mode whose amplitude grows in time, i.e., an unstable mode. Solutions with $\gamma = 0$ were called {\em neutrally stable solutions} by \citet{bodo1,bodo2} and correspond to trapped waves \citep{edwinroberts}. In such a case, solutions with $k_{\rm i}^2 > 0$ are body waves and those with $k_{\rm i}^2 < 0$ are surface waves \citep[which were respectively called {\em reflected} and {\em ordinary} modes by][]{bodo1,bodo2}. In the general case, $\gamma \neq 0$ and $k_{\rm i}^2$ is complex, meaning that modes have mixed properties since no pure body or surface modes are possible. If $\Re \left( k_{\rm i}^2 \right) > \Im \left( k_{\rm i}^2 \right) $, as expected for a small velocity shear, the dominant internal wave character is determined by the sign of $\Re \left( k_{\rm i}^2 \right)$. Regarding the behavior of solutions in the external medium, we must note that $k_{\rm e}^2$ is also a complex quantity for complex $\omega$, so wave modes have mixed propagating and evanescent properties. For this reason, we use the more convenient representation of the external solution in terms of Hankel functions instead of the modified Bessel functions usually considered for trapped waves with real $\omega$ \citep[see][]{cally}. It is straight-forward to see from Equation~(\ref{eq:ke}) that when $\omega$ is real and $\omega^2 < k_z^2 \vae^2$, $k_{\rm e}^2$ is real and negative. In such a case, one can directly express the function $H_m^{(1)} \left( k_{\rm e} a \right)$ in terms of $K_m \left( k_{\rm e} a \right)$ \citep{abramowitz}.

In the following investigation, we restrict ourselves to modes with $m \neq 0$. This is done so because the azimuthal flow has no effect on $m=0$ modes in the linear regime. In addition, it is unlikely that the development of small spatial-scales at the cylinder boundary obtained by \citet[see their Figure~1]{terradas} are related to sausage-like ($m=0$) solutions but more likely to modes with large $m$, and therefore, with small azimuthal length-scales. Hence, the reader is referred to \citet{bodo1} for a study of the $m = 0$ case. In addition, we only consider modes which are radially fundamental.

\subsection{Thin Tube Approximation}

To shed more light on the nature of solutions of Equation~(\ref{eq:reldisper}), we consider the thin tube (TT) approximation, i.e., $k_z a \ll 1$, which is realistic in the context of transversely oscillating coronal loops. Then, we perform a first order expansion for small arguments of the Bessel and Hankel functions in Equation~(\ref{eq:reldisper}) in the case $m \neq 0$. By this procedure, we obtain the TT version of the dispersion relation, namely
\begin{equation}
 \rho_{\rm i}  \left[ \left( \Omega_{\rm i}^2 - k_z^2 \vai^2 \right)^2 - 4 v_0^2 \Omega_{\rm i}^2 \right] + \rho_{\rm e}  \left( \omega^2 - k_z^2 \vae^2 \right) \left[ \left( \Omega_{\rm i}^2 - k_z^2 \vai^2 \right) - 2 v_0 \Omega_{\rm i} \frac{m}{\left|  m \right|}  \right] = 0. \label{eq:ttdisper} 
\end{equation}
For a very small velocity shear, one can roughly neglect the terms $4 v_0^2 \Omega_{\rm i}^2$ and $2 v_0 \Omega_{\rm i} m/\left|  m \right|$, so the effect of the shear flow is only kept in $\Omega_{\rm i}$. In such a case, Equation~(\ref{eq:ttdisper}) becomes a second-order polynomial in $\omega$, whose analytical solution is
\begin{equation}
 \omega \approx \frac{\rho_{\rm i}}{\rho_{\rm i} + \rho_{\rm e}} m v_0 \pm \left[ \frac{\left( \rho_{\rm i} \vai^2 + \rho_{\rm e} \vae^2\right)}{\left( \rho_{\rm i} + \rho_{\rm e} \right)} k_z^2 - \frac{\rho_{\rm i} \rho_{\rm e}}{\left( \rho_{\rm i} + \rho_{\rm e} \right)^2} m^2 v_0^2 \right]^{1/2}. \label{eq:approxw}
\end{equation}
This solution corresponds to the usual forward ($\omega_{\rm R} > 0$) and backward ($\omega_{\rm R} < 0$) body waves modified by the shear flow. We see that the KHI takes place for a velocity shear at the cylinder boundary that satisfies
\begin{equation}
\left( \frac{v_0 a}{\vai} \right) > \sqrt{2 \left( \frac{\rho_{\rm i}}{\rho_{\rm e}} + 1 \right)}\,  \frac{k_z a }{\left| m \right|}. \label{eq:critical}
\end{equation}
For a velocity shear larger than the critical value, a damped mode ($\gamma < 0$) and an unstable mode ($\gamma > 0$) are present. We focus on the unstable solution, whose growth rate in the TT case can be easily obtained from Equation~(\ref{eq:approxw}), namely
\begin{equation}
 \gamma =  \left[ \frac{\rho_{\rm i} \rho_{\rm e}}{\left( \rho_{\rm i} + \rho_{\rm e} \right)^2} m^2 v_0^2  -  \frac{\left( \rho_{\rm i} \vai^2 + \rho_{\rm e} \vae^2\right)}{\left( \rho_{\rm i} + \rho_{\rm e} \right)} k_z^2\right]^{1/2}. \label{eq:gamma}
\end{equation}
Note that Equation~(\ref{eq:gamma}) is formally identical to Equation~(1) of \citet{terradas}, although they obtained their expression by assuming incompressible perturbations in a discontinuous interface. One has to bear in mind that Equations~(\ref{eq:critical}) and (\ref{eq:gamma}) are only approximately valid for a very small velocity shear and miss some effects such as, for example, the dependence on the sign of $m$. Therefore, both the actual critical velocity shear for destabilization and the growth rate might be significantly different. However, Equations~(\ref{eq:critical}) and (\ref{eq:gamma}) allow us to predict two important results. First, the longitudinal wavenumber has a stabilizing effect since the larger $k_z$, the larger the critical velocity shear and the smaller $\gamma$; and second, modes with large $m$ are more unstable than those with small $m$.  This conclusion is equivalent to the well-known general result of the KHI in magnetic interfaces \citep[e.g.,][]{chandra,drazin} concerning the stabilizing role of the wavenumber component along the magnetic field direction.

\section{RESULTS}
\label{sec:results}

\subsection{Instability Regimes}

\begin{figure}[!htp]
\centering
 \epsscale{0.5}
 \plotone{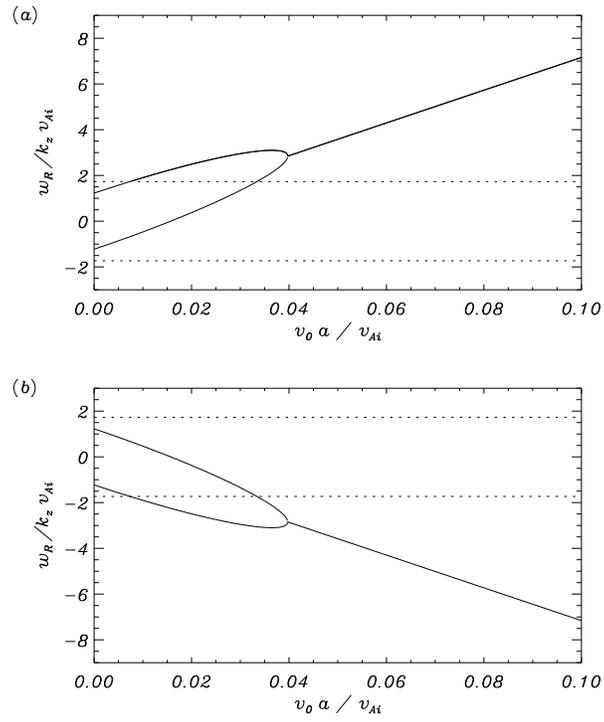}
\caption{Normalized phase velocity versus the velocity shear at the cylinder boundary for $\rho_{\rm i} / \rho_{\rm e} = 3$, $L/a = 100$, and $n=1$. (a) Solution with $m=4$. (b) Solution with $m=-4$. Dotted lines denote the external cut-off phase velocities, i.e., $\omega/k_z = \pm \vae$. \label{fig:solutions}}
\end{figure}

First, we study the critical velocity shear that leads to the instability of the solutions of the dispersion relation (Equation~(\ref{eq:reldisper})). Unless otherwise stated, we consider the next set of parameters: $\rho_{\rm i} / \rho_{\rm e} = 3$, $L/a = 100$, and $n=1$. We restrict ourselves to a velocity shear much smaller than the internal Alfv\'en speed, i.e., $v_0 a / \vai < 0.1$. Equation~(\ref{eq:reldisper}) has been solved for different values of the azimuthal wavenumber $m$. As expected, the degeneracy between forward ($\omega_{\rm R} > 0$) and backward ($\omega_{\rm R} < 0$) waves is broken by the presence of flow. When the critical value of the velocity shear is reached, the phase velocity of both forward and backward waves coincides and the KHI appears. We find that, in the range of considered velocity shears, the first unstable modes are those with $m = \pm 4$, and Figure~\ref{fig:solutions} shows their phase velocity. The behavior of other solutions with larger $m$ is similar in all cases, the critical velocity shear being shifted towards smaller values as $|m|$ increases. Note that solutions become leaky, i.e., their phase velocity exceeds the external cut-off values, denoted by horizontal dotted lines in Figure~\ref{fig:solutions}, before the KHI takes place. 

\begin{figure}[!htp]
\centering
 \epsscale{0.5}
 \plotone{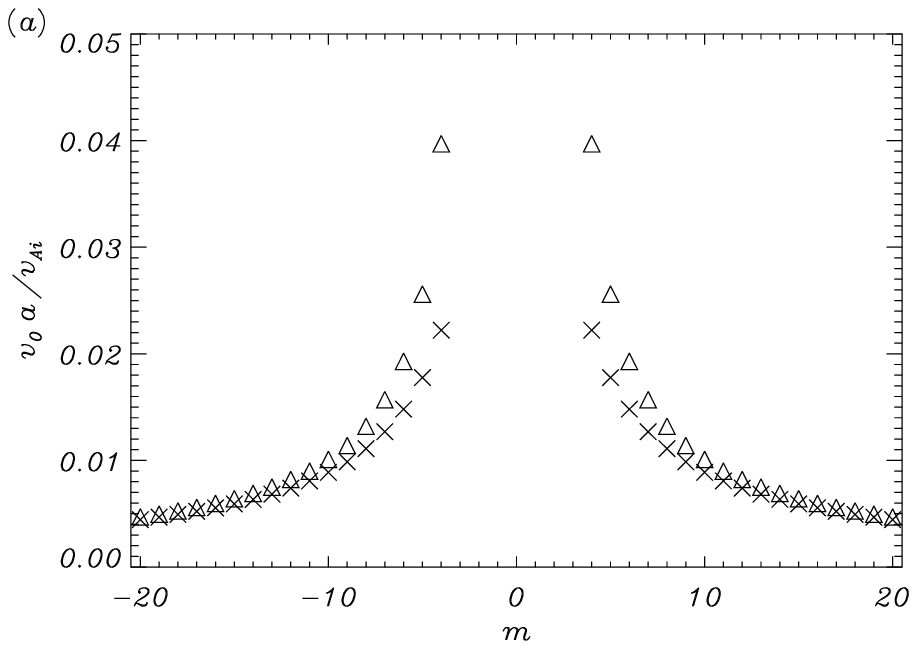}
 \plotone{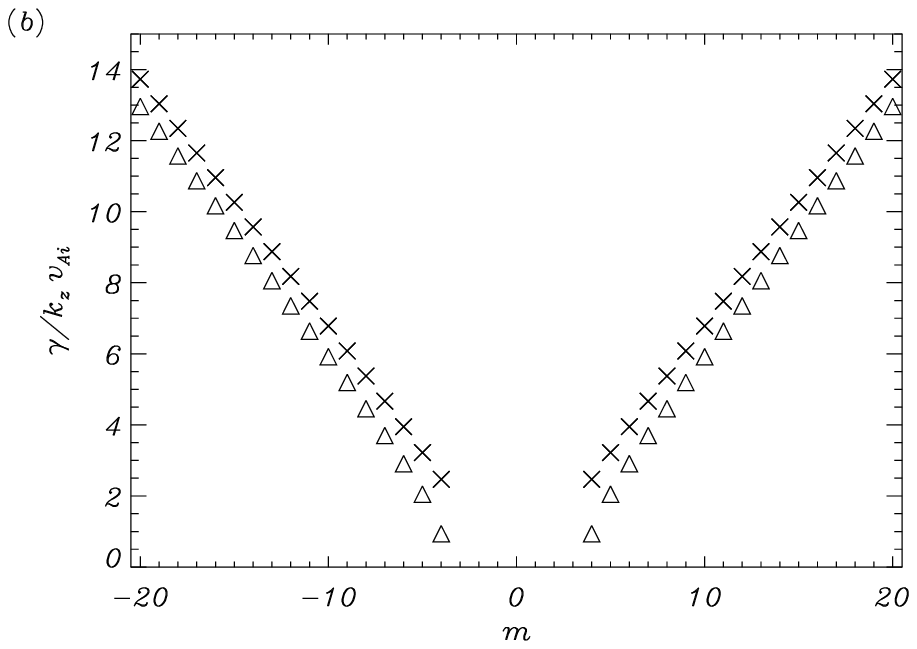}
\caption{(a) Critical velocity shear at the tube boundary as a function of $m$. The symbol $\triangle$ denotes the actual value obtained by solving the dispersion relation, whereas the symbol $\times$ corresponds to the approximation given by Equation~(\ref{eq:critical}).  (b) Normalized growth rate as a function of $m$ for $v_0 a / \vai = 0.05$. Again, the symbol $\triangle$ denotes the actual solution and the symbol $\times$ corresponds to the approximation given by Equation~(\ref{eq:gamma}). In all cases, $\rho_{\rm i} / \rho_{\rm e} = 3$, $L/a = 100$, and $n=1$ have been considered. \label{fig:comp}}
\end{figure}

We compare in Figure~\ref{fig:comp}(a) the actual critical velocity shear with that obtained from the approximation given by Equation~(\ref{eq:critical}). For small $|m|$, the actual critical velocity is larger than the approximated value. For $|m|=4$ there is a factor 2, approximately, between the actual value and the approximation. The agreement between both values is substantially better as $|m|$ increases. On the other hand, Figure~\ref{fig:comp}(b) displays the growth rate of the unstable modes with $-20 \leq m \leq 20$ for $v_0 a / \vai = 0.05$. The actual growth rate is compared to that predicted by Equation~(\ref{eq:gamma}) in the TT case. We see that the approximated value is slightly larger than the actual one. Again, the difference between both of them gets smaller as the azimuthal wavenumber increases. It is worth noting that results for positive and negative values of $m$ do not show significant differences regarding both the critical velocity and the growth rate. The reason for this behavior is that the considered velocity shears are very small, thus the degeneracy between positive and negative values of $m$ is only slightly broken. For larger velocity shear, a significant dependence on the sign of $m$ appears.

\subsection{Parametric Study}

In this section we study the dependence of the KHI growth rate, $\gamma$, on different parameters. As stated before, since for the considered shear flow velocities there are no significant differences between solutions with positive and negative values of $m$, hereafter we only show the results corresponding to positive values of $m$.

\begin{figure}[!htp]
\centering
 \epsscale{0.5}
 \plotone{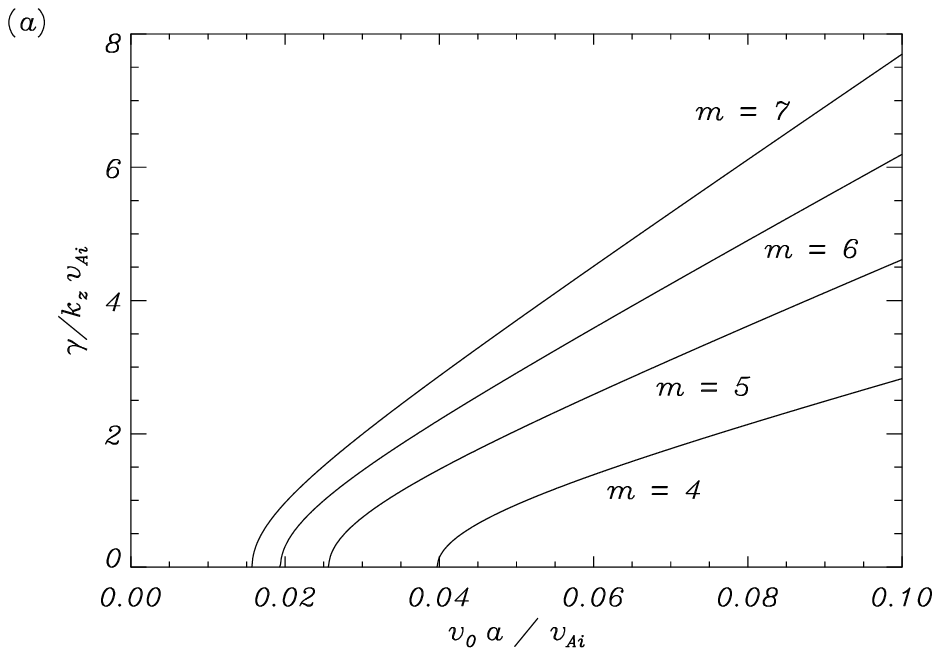}
 \plotone{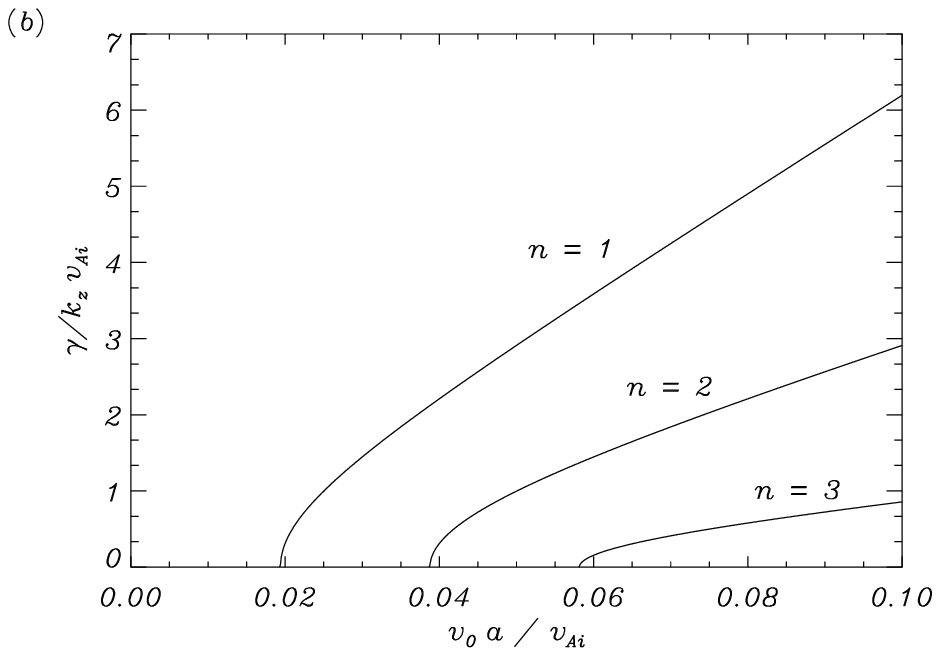}
 \plotone{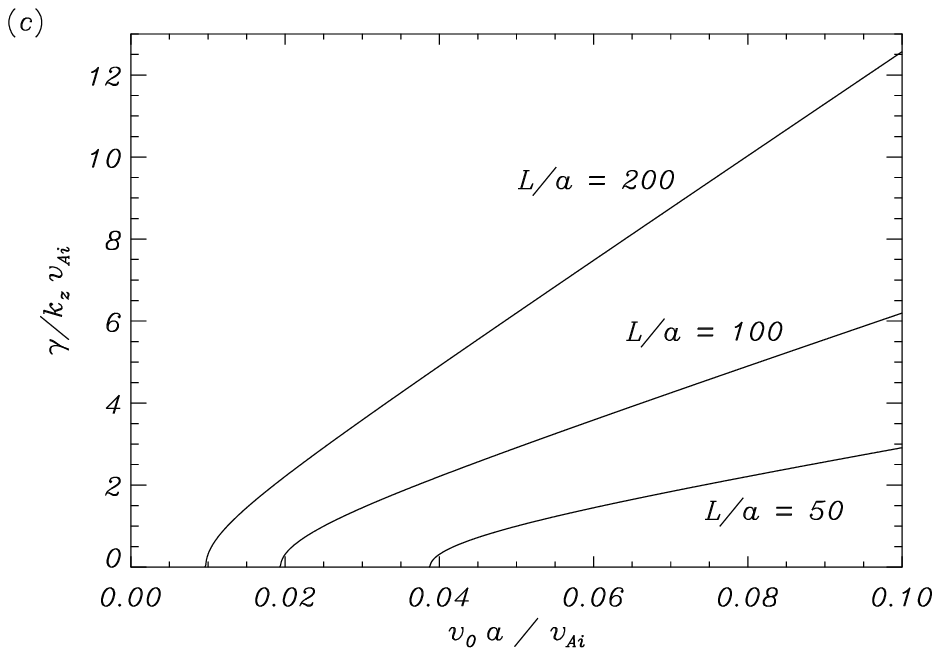}
\caption{Normalized growth rate as a function of the velocity shear at the cylinder boundary. (a) Results for $m=$~4, 5, 6, and 7 with $\rho_{\rm i} / \rho_{\rm e} = 3$, $L/a = 100$, and $n=1$. (b) Results for $n=$~1, 2, and 3 with $\rho_{\rm i} / \rho_{\rm e} = 3$, $L/a = 100$, and $m=6$. (c) Results for $L/a=$~50, 100, and 200 with $\rho_{\rm i} / \rho_{\rm e} = 3$, $m = 6$, and $n=1$.\label{fig:param}}
\end{figure}

First, we consider $\rho_{\rm i} / \rho_{\rm e} = 3$, $L/a = 100$, and $n=1$, and study the dependence of $\gamma$ with the velocity shear for different values of the azimuthal wavenumber, $m$. These results are plotted in Figure~\ref{fig:param}(a). As was already commented, we obtain that the larger $m$, the larger $\gamma$ and the smaller the critical velocity shear. Next, we investigate the dependence of $\gamma$ on the longitudinal wavenumber, $k_z$. We have fixed $m=6$ in the following computations, although equivalent results are obtained when other values of $m$ are considered. The value of $k_z$ depends on both $n$ and $L$ through Equation~(\ref{eq:kz}). Figure~\ref{fig:param}(b) displays the results for different values of $n$.  We see that as the value of $n$ increases, i.e., $k_z$ grows, the critical velocity shear is shifted toward larger values and the growth rate is reduced. This behavior qualitatively agrees with the analytical predictions of the TT case (see Equations~(\ref{eq:critical}) and (\ref{eq:gamma})), and indicates that longitudinal harmonics are more stable than the corresponding fundamental mode. On the other hand, Figure~\ref{fig:param}(c) shows the results for different values of $L/a$. As $L/a$ increases, $k_z$ decreases. We therefore obtain that long loops are more unstable than short loops for fixed values of $m$ and $n$. 

\begin{figure}[!htp]
\centering
 \epsscale{0.5}
 \plotone{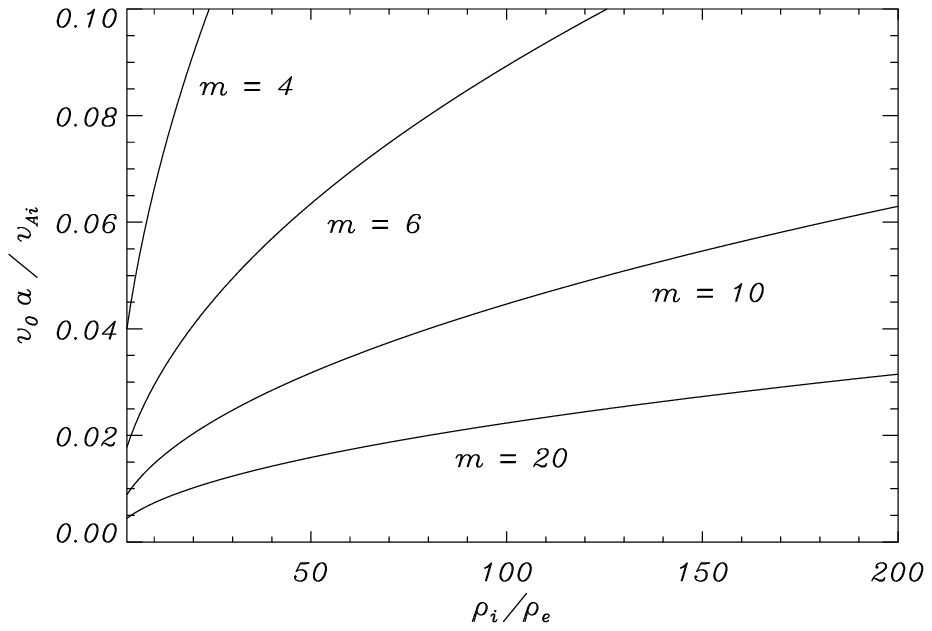}
\caption{Critical velocity shear at the tube boundary as a function of $\rho_{\rm i} / \rho_{\rm e}$ for $m=$~4, 6, 10, and 20. In all calculations, $L/a = 100$ and $n = 1$. \label{fig:param2}}
\end{figure}

Finally, Figure~\ref{fig:param2} displays the critical velocity shear as a function of the density contrast, $\rho_{\rm i} / \rho_{\rm e}$. Whereas small values of $\rho_{\rm i} / \rho_{\rm e}$ are realistic in the context of transversely oscillating coronal loops, larger values ($\rho_{\rm i} / \rho_{\rm e} \sim200$) are usually considered to model filament and prominence fine-structures (threads). Since kink waves have been detected in filament threads \citep[e.g.,][]{okamoto,hinode,lin}, it seems appropriate to extend our present study to the context of prominences. We obtain that the critical velocity shear of modes with small $m$ increases dramatically as the density contrast grows, although this effect is not so important when large values of $m$ are taken into account. This dependence on the density contrast can be seen in Equation~(\ref{eq:critical}). These results suggest that the KHI is much more difficult to develop in filament thread conditions in comparison to coronal loop conditions, meaning that filament threads might be more stable than coronal loops when shear flows are present. 

\subsection{Energy considerations}

Results from previous Sections indicate that the larger the azimuthal wavenumber $m$, the more unstable the mode. However, we also have to bear in mind that modes with large $m$ could be very difficult to excite by an external disturbance, say a flare, and therefore would weakly contribute to the development of the KHI. \citet{terradasenergy} studied how the energy from an initial disturbance is distributed in different eigenmodes of a flux tube. By considering the mathematical method by \citet{rudermanroberts06}, these authors analytically computed the amplitude and the amount of energy deposited in each eigenmode. Although this calculation depends on the form of the initial disturbance, \citet{terradasenergy} concluded that the trapped energy of the eigenmode dramatically decreases as the azimuthal wavenumber $m$ is increased. Typically, the difference in energy trapped by modes with consecutive $m$ is around three orders of magnitude \citep[see Figure~7(b) of][]{terradasenergy}. Let us assume that the trapped energy by the kink ($m=1$) mode is $E_1$. Then, the energy initially deposited in the eigenmode with azimuthal wavenumber $m$, namely $E_m \left(t=0\right)$, can be roughly estimated as
\begin{equation}
 E_m \left(t=0\right) \approx E_1 \times 10^{-3 \left( \left| m \right| -1 \right)}. \label{eq:ene0}
\end{equation}
The reader must be aware that Equation~(\ref{eq:ene0}) is a highly approximated expression since the actual deposited energy depends strongly on the form of the initial disturbance among other effects. So, the following calculations should be interpreted from a qualitatively point of view. 

On the other hand and according to \citet{bray}, the wave energy density, $e$, is quadratic in the perturbations, thus
\begin{equation}
 e = \frac{1}{2} \left[ \rho_0 \left( v_r^2 + v_\varphi^2 \right) + \frac{1}{\mu} \left( B_r^2 + B_\varphi^2 + B_z^2 \right) \right].
\end{equation}
If a mode is unstable, its related energy density will increase in time by a factor $\exp \left( 2 \gamma_m t \right)$, where $\gamma_m$ is the growth rate for the azimuthal wavenumber $m$. Therefore, by adding this factor to Equation~(\ref{eq:ene0}), we get a very approximated estimation for the time-dependent energy of an unstable mode with respect to the kink mode energy,
 \begin{equation}
E_m \left(t\right)  = E_m \left(t=0\right) \times  \exp \left( 2 \gamma_m t \right) \approx E_1 \times 10^{-3 \left( \left| m \right|  -1 \right)} \exp \left( 2 \gamma_m t \right). \label{eq:enetime}
\end{equation}
The quantity $E_m\left(t\right)  / E_1$ for a fixed time indicates the effective instability of an eigenmode, since it takes into account both the initial energy deposited in the mode during its excitation, and its subsequent increase due to the instability. A comparison of the value of $E_m \left(t\right) / E_1$ corresponding to different modes indicates which of them 
are the more effectively unstable a the given time-scale, $t$. For the present application, an appropriate time-scale is $t =P_k / 2 $, where  $P_k$  is the kink mode period. Half a kink mode period is the maximum time for the KHI to develop before shear flows change direction due to the flux tube kink motion, and so our assumption that the flow is time-independent is not fulfilled.

\begin{figure}[!htp]
\centering
\epsscale{0.5}
\plotone{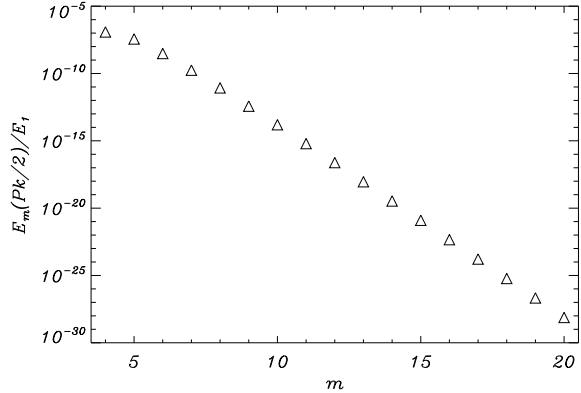}
\caption{$E_m \left(P_k / 2\right) / E_1$ of the first unstable modes, considering  $\rho_i / \rho_e = 3$, $L/a = 100$, and $n=1$. \label{fig:energy}}
\end{figure}

Figure~\ref{fig:energy} displays $E_m \left(P_k / 2\right) / E_1$ for the modes with $4 \leq m \leq 20$.  We see that although the growth rate is larger for modes with large $m$, they cannot acquire enough energy during half a kink period to become more relevant than solutions with small $m$. This qualitative analysis points out that only the first two or three unstable modes could effectively contribute to the development of a KHI during the kink motion of the flux tube. However, since $E_m \left(P_k / 2\right) / E_1$ takes very small values, e.g., $E_m \left(P_k / 2\right) / E_1 \approx 10^{-7}$ for the most effectively unstable mode, the KHI cannot remove enough energy from the flow to be relevant for the damping of kink oscillations, at least in the linear regime.

\subsection{Stabilization by Magnetic Twist}

Results from previous sections indicate that a KHI should develop during kink oscillations of coronal loops. However,  as these kind of instabilities have not been reported to date, it is likely that some mechanism, not considered here, is able to suppress the KHI in coronal loops. From studies of the KHI in simple configurations \citep[e.g.,][]{chandra}, it is known that the inclusion of a component of the magnetic field along the flow direction has a stabilizing effect. In our cylindrical configuration, this corresponds to an azimuthal component of the magnetic field, so field lines would be twisted. The investigation of the effect of magnetic twist on the MHD eigenmodes of a flux tube is mathematically complicated and has been broached by a number of authors using several approximations. For example, \citet{bogdan}, \citet{bennet}, and \citet{fedun1} studied wave propagation in twisted tubes by taking the incompressible approximation into account. \citet{carter1} and \citet{carter2,carter3} investigated wave modes in cylinders with the magnetic twist being constrained within an annulus,  whereas \citet{fedun2} only considered $m=0$ modes in fully twisted tubes. Finally, \citet{ruderman} studied non-axisymmetric oscillations of twisted tubes in the thin tube approximation. It is beyond the purpose and scope of the present work to include magnetic twist in our cylindrical configuration, since it would complicate matters, e.g., by introducing resonances in the system. Instead, we consider in this section the more simple Cartesian (slab) geometry, and study how a component of the magnetic field along the flow direction affects the KHI. An analogy between the slab case and the cylinder case can be performed. 

Let us consider a magnetic slab with half-width $a$ in the $x$-direction embedded in an unlimited environment. Both the $y$- and $z$-directions are unbounded. All symbols in the following equations have the same meaning as in previous sections. We assume a mass flow in the $y$-direction, $ \mathbf{U}_0 = U_0 \hat{e}_y$. The following density, $\rho_0$, and flow velocity, $U_0$, profiles are assumed,
\begin{equation}
 \rho_0 = \left\{ \begin{array}{lcl}
                         \rho_{\rm i}, & \textrm{if} &  | x | \leq a, \\
			\rho_{\rm e}, & \textrm{if} &  | x | > a.
                        \end{array}
  \right., \qquad   U_0 = \left\{ \begin{array}{lcl}
                         v_0 a, & \textrm{if} &  | x | \leq a, \\
			0, & \textrm{if} &  | x | > a.
                        \end{array}
  \right.
\end{equation}
The magnetic field is
\begin{equation}
 \mathbf{B}_0= \left\{ \begin{array}{lcl}
                         B_0 \cos \alpha\, \hat{e}_y + B_0 \sin \alpha\, \hat{e}_z, & \textrm{if} &  | x | \leq a, \\
			B_0 \hat{e}_z, & \textrm{if} &  | x | > a.
                        \end{array}
  \right.
\end{equation}
Thus, for $\alpha = \pi / 2$ the internal magnetic field is perpendicular to the flow, whereas for $\alpha = 0$ the internal magnetic field and the flow are in the same direction. We now apply Equations~(\ref{eq:momentum})--(\ref{eq:induction}) to this configuration and assume perturbations proportional to $\exp \left( i k_y y+ i k_z z - i \omega t \right)$, with $k_z$ given by Equation~(\ref{eq:kz}) and $k_y = m/a$. This form of $k_y$ has been assumed in a number of works \citep[e.g.,][]{hollwegyang,arregui} in order to extend the results of the slab models to the cylindrical geometry. In particular, \citet{arregui} showed that the solutions of the slab model tend to those of the cylindrical configuration when larger values of $k_y$ in comparison with $k_z$ are taken into account. In the present case, $k_y \gg k_z$ as long as $L/a \gg n \pi / m$.

The general dispersion relation of MHD modes supported by the slab model is given in Equation~(7) of \citet{joarder}. In our present notation and for $\beta = 0$, the dispersion relation becomes,
\begin{equation}
 \rho_{\rm i} \mathcal{K}_{\rm e} \left[ \Omega_{\rm i}^2 - k^2 \vai^2 \sin^2 \left( \theta + \alpha \right) \right] \left\{ \begin{array}{c} \tanh \left( \mathcal{K}_{\rm i} a \right) \\ \coth \left( \mathcal{K}_{\rm i} a \right)  \end{array} \right\}  + \rho_{\rm e} \mathcal{K}_{\rm i} \left( \omega^2 - k_z^2 \vae^2  \right) = 0, \label{eq:slabdisper}
\end{equation}
with 
\begin{equation}
 \mathcal{K}_{\rm i}^2 = \frac{k^2 \vai^2 - \Omega^2}{\vai^2}, \qquad  \mathcal{K}_{\rm e}^2 = \frac{k^2 \vae^2 - \omega^2}{\vae^2}, 
\end{equation}
where $\Omega_{\rm i} = \omega - m v_0$, $k^2 = k_y^2 + k_z^2$, and $\theta = \arctan \left( k_y /k_z\right)$. The $\tanh$ term in Equation~(\ref{eq:slabdisper}) corresponds to symmetric solutions whereas the $\coth$ term stands for anti-symmetric solutions with respect to the $x$-axis. In the case $\alpha = \pi / 2$ and $v_0 = 0$, Equation~(\ref{eq:slabdisper}) reduces to the dispersion relation derived by \citet{raeroberts}.

\begin{figure}[!htp]
\centering
 \epsscale{0.5}
 \plotone{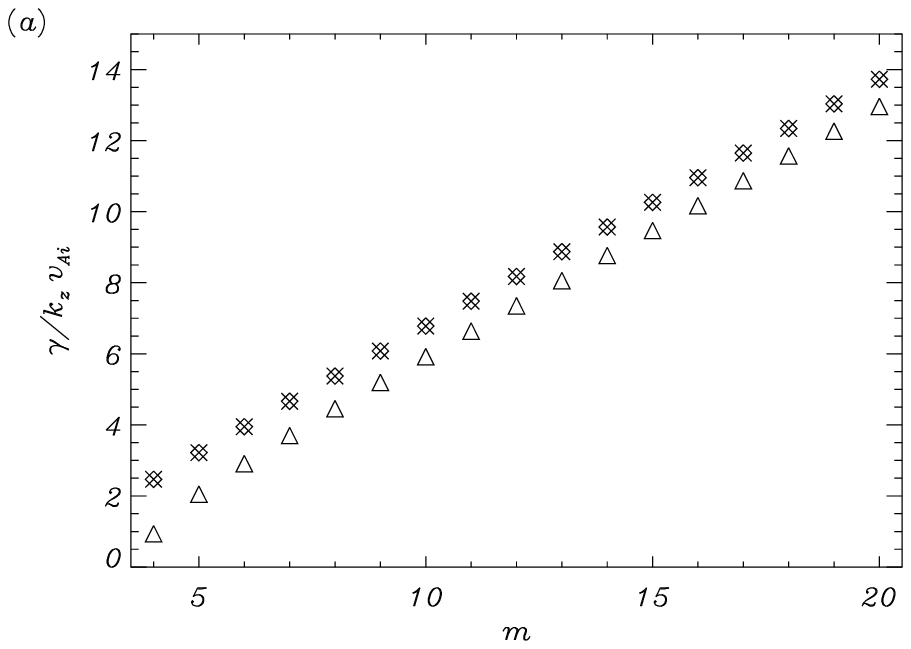}
\plotone{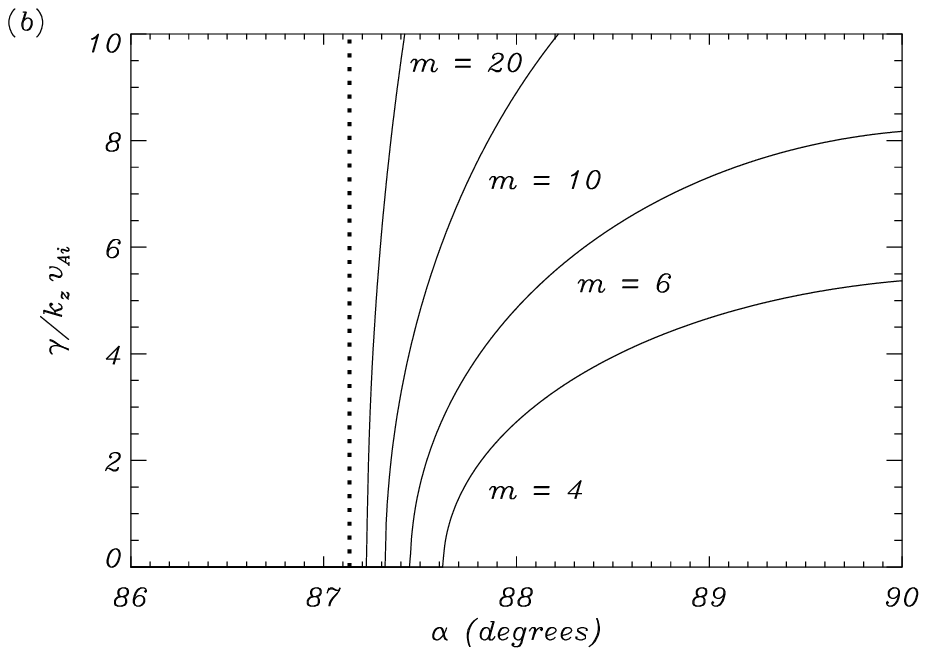}
\caption{(a) Normalized growth rate as a function of $m$ for $v_0 a / \vai = 0.05$. The symbol $\triangle$ denotes the solution of the cylinder case, whereas symbols $\times$ and $\Diamond$ correspond to the symmetric and anti-symmetric slab solutions, respectively.  (b) Normalized growth rate of symmetric slab solutions as a function of $\alpha$ for $m=$~4, 6, 10, and 20 with $v_0 a / \vai = 0.1$. The vertical dotted line denotes the critical angle for $m \to \infty$ given by Equation~(\ref{eq:critalpha}). In all cases, $\rho_{\rm i} / \rho_{\rm e} = 3$, $L/a = 100$, and $n=1$ have been considered. \label{fig:slab}}
\end{figure}

Figure~\ref{fig:slab}(a) shows a comparison of the previously obtained growth rates in the cylinder case with those computed in the slab case by solving Equation~(\ref{eq:slabdisper}). We see that the growth rates are slightly larger in the slab case than in the cylinder case but both solutions are in reasonable agreement, the difference being reduced as $m$ increases. This behavior coincides with the results of \citet{arregui} for the real part of the frequency. On the other hand,  symmetric and anti-symmetric modes in the slab case have very similar growth rates, so hereafter we restrict ourselves to symmetric modes for simplicity. Next, Figure~\ref{fig:slab}(b) displays the growth rate of symmetric modes as a function of $\alpha$ for $v_0 a / \vai = 0.1$ . We see that, depending on $m$, there is a critical $\alpha$ which suppresses the KHI. The critical  $\alpha$ clusters towards a limit value as $m$ increases. To obtain an analytical expression for this limit value of $\alpha$, let us consider the situation $m \gg k_z a$. In such a case, the length-scale in the $y$-direction is much smaller than those in the $x$- and $z$-directions, and the solutions of the slab configuration can be approximated by those of a single interface. In addition, if we assume incompressible perturbations, the dispersion relation is \citep{joardersatya,joardernaka},
\begin{equation}
 \rho_i  \left( \Omega^2 - k^2 \vai^2 \sin^2 \left( \theta + \alpha \right) \right) +  \rho_e \left( \omega^2 - k_z^2 \vae^2  \right) = 0,\label{eq:inter}
\end{equation}
whose exact analytical solution is
\begin{equation}
 \omega = \frac{\rho_{\rm i}}{\rho_{\rm i} + \rho_{\rm e}} m v_0  \pm \left[ \frac{\left(\rho_{\rm i} \vai^2 k^2\sin^2 \left( \theta + \alpha \right)+ \rho_{\rm e} \vae^2 k_z^2\right)}{\left(\rho_{\rm i} + \rho_{\rm e} \right)}  - \frac{\rho_{\rm i} \rho_{\rm e}}{\left(\rho_{\rm i} + \rho_{\rm e} \right)^2} m^2 v_0^2 \right]^{1/2}. \label{eq:interex}
\end{equation}
Note the remarkable resemblance between Equations~(\ref{eq:approxw}) and (\ref{eq:interex}), although the former was obtained for a cylinder in the TT and $\beta=0$ approximations, and the latter corresponds to a magnetic interface in the incompressible limit. In fact, it is possible to directly obtain Equation~(\ref{eq:interex}) from Equation~(\ref{eq:approxw}) by simply adding the contribution of an azimuthal component of the magnetic field considering its value at the cylinder edge. Nevertheless, we have derived Equation~(\ref{eq:interex}) from the expressions of the Cartesian case for the sake of a more rigorous method. 

Finally, it is straight-forward to derive from Equation~(\ref{eq:interex}) the condition for stability,
\begin{equation}
 \sin^2 \left( \theta + \alpha \right) = \frac{\rho_{\rm e}}{\left(\rho_{\rm i}  + \rho_{\rm e}  \right)} \frac{m^2}{\left(m^2 + k_z^2 a^2\right)} \left( \frac{v_0 a}{\vai}  \right)^2 - \frac{k_z^2 a^2}{m^2 + k_z^2 a^2}.
\end{equation}
By performing the limit $m \gg k_z a$ one obtains that the critical $\alpha$ satisfies,
\begin{equation}
 \cos \alpha \approx \sqrt{\frac{\rho_e}{\left(\rho_i + \rho_e\right)}} \left( \frac{v_0 a}{\vai}  \right). \label{eq:critalpha}
\end{equation}
The value obtained from Equation~(\ref{eq:critalpha}) is indicated in Figure~\ref{fig:slab}(b) by a vertical line, showing a good agreement with the results. For $\alpha$ smaller than the limit critical value, the KHI is completely suppressed for any value of $m$. For $v_0 a / \vai = 0.1$, the limit critical angle is $\alpha \approx 87\deg$, which corresponds to a small inclination of the magnetic field with respect to the $z$-axis, i.e., a very small azimuthal component of the magnetic field in the analog cylindrical case. So, a very ``weak'' twist of magnetic field lines can be enough to prevent the triggering of the KHI in a cylindrical coronal loop, the azimuthal component of the magnetic field needed to suppress the KHI being probably much smaller than the critical value of the pinch instability. A weak twist of magnetic field lines is very likely and realistic in the context of coronal loops. Although further studies are needed, magnetic twist seems to be a consistent explanation for the absence of the KHI in the observations of transversely oscillating loops. 

\section{CONCLUSION}
\label{sec:conclusion}

In this paper, we have studied the linear KHI in straight, untwisted, and cylindrical coronal magnetic flux tubes due to the presence of azimuthal shear flows at the tube boundary. We have found that for realistic values of the velocity shear, fluting-like MHD modes with large azimuthal wavenumber and small longitudinal wavenumber are unstable. A qualitative analysis based on energy considerations suggests that only the first unstable fluting modes can develop a KHI during time-scales related to kink motions of the flux tube. Since these kind of instabilities have not been reported by observers to date, we considered an analog in Cartesian geometry and obtained that the twist of magnetic field lines could completely suppress the KHI in twisted coronal flux tubes. The magnetic twist that can suppress the KHI is very weak and probably much smaller than the critical twist of the pinch instability, meaning that the magnetic stability of the flux tube would not be compromised. Further investigations on this issue are needed.

Some effects not considered here could be studied in future works. The next logical step is to assume a time and azimuthal dependence of the flow in order to describe more realistically the actual motions of a kink oscillation. This investigation should be numerically performed. Other effects are, for example, to consider the $\beta \neq 0$ case, or to take into account the presence of a transitional layer between the loop and the corona.

\acknowledgements{
    RS thanks the Conselleria d'Innovaci\'o, Interior i Just\'icia of the CAIB for a fellowship. RS, RO, and JLB acknowledge the financial support received from the Spanish MICINN, FEDER funds, and the CAIB under Grants No. AYA2006-07637 and PCTIB-2005GC3-03.  JT also acknowledges support form K.U. Leuven via GOA/2009-009.}


\begin{thebibliography}{}
%
\bibitem[Abramowitz \& Stegun(1972)]{abramowitz} Abramowitz, M., \& Stegun, I. A. 1972, Handbook of Mathematical Functions (New York: Dover Publications)

\bibitem[Andries et al.(2000)]{andries1} Andries, J., Tirry, W. J., \& Goossens, M. 2000, \apj, 531, 561

\bibitem[Andries \& Goossens(2001)]{andries2} Andries, J., \& Goossens, M. 2001, \aap, 368, 1083

\bibitem[Arregui et al.(2007)]{arregui} Arregui, I., Terradas, J., Oliver, R., \& Ballester, J. L. 2007, \solphys, 246, 213

\bibitem[Aschwanden et al.(1999)]{asch} Aschwanden, M. J., Fletcher, L, Schrijver, C. J., \& Alexander, D. 1999, \apj, 520, 880

\bibitem[Bennet et al.(1999)]{bennet} Bennet, K., Roberts, B., \& Narain, U. 1999, \solphys, 185, 41

\bibitem[Bodo et al.(1989)]{bodo1} Bodo, G., Rosner, R., Ferrari, A., \& Knobloch, E. 1989, \apj, 341, 631
\bibitem[Bodo et al.(1996)]{bodo2} Bodo, G., Rosner, R., Ferrari, A., \& Knobloch, E. 1996, \apj, 470, 797

\bibitem[Bogdan(1984)]{bogdan} Bogdan, T. J. 1984, \apj, 282, 769

\bibitem[Bondeson et al.(1987)]{bondeson} Bondeson, A., Iacono, R., \& Bhattacharjee, A. 1987, Phys. Fluids, 30, 2167

\bibitem[Bray \& Loughhead(1974)]{bray} Bray, R. J., \& Loughhead, R. 1974, The Solar Chromosphere (ed. Chapmam)

\bibitem[Cally(1986)]{cally} Cally, P. S. 1986, \solphys, 103, 277


\bibitem[Carter \& Erd\'elyi(2007)]{carter2} Carter, B. K., \& Erd\'elyi, R. 2007, \aap, 475, 323

\bibitem[Carter \& Erd\'elyi(2008)]{carter3} Carter, B. K., \& Erd\'elyi, R. 2008, \aap, 481, 239

\bibitem[Chandrasekhar(1961)]{chandra} Chandrasekhar, S. 1961, Hydrodynamic and Hydromagnetic Stability (Oxford Clarendon Press, London)

\bibitem[Drazin \& Reid(1981)]{drazin} Drazin, P., \& Reid, W. 1981, Hydrodynamic Stability (Cambridge University Press, New York)

\bibitem[Edwin \& Roberts(1983)]{edwinroberts} Edwin, P. M., \& Roberts, B. 1983, \solphys, 88, 179

\bibitem[Erd\'elyi et al.(1995)]{robertusflow} Erd\'elyi, R., Goossens, M., \& Ruderman, M. S. 1995, \solphys, 161, 123

\bibitem[Erd\'elyi \& Fedun(2006)]{fedun1} Erd\'elyi, R., \& Fedun, V. 2006, \solphys, 238, 41

\bibitem[Erd\'elyi \& Fedun(2007)]{fedun2} Erd\'elyi, R., \& Fedun, V. 2007, \solphys, 246, 101

\bibitem[Erd\'elyi \& Carter(2006)]{carter1} Erd\'elyi, R., \& Carter, B. K. 2006, \aap, 455, 361

\bibitem[Goossens et al.(1992)]{goossens92} Goossens, M., Hollweg, J. V., \& Sakurai, T. 1992, \solphys, 138, 233

\bibitem[Goossens et al.(2002)]{goossens02} Goossens, M., Andries, J., \& Aschwanden, M. J. 2002, \aap, 394, L39

\bibitem[Goossens et al.(2009)]{goossens} Goossens, M., Terradas, J., Andries, J., Arregui, I., Ballester, J. L. 2009, \aap, 503, 213

\bibitem[Hollweg \& Yang(1988)]{hollwegyang} Hollweg, J. V., \& Yang, J. 1988, \jgr, 93, 5423

\bibitem[Holzwarth et al.(2007)]{holzwarth} Holzwarth, V., Schmitt, D., \& Sch\"ussler, M. 2007, \aap, 469, 11

\bibitem[Joarder(2002)]{joarder} Joarder, P. S. 2002, \aap, 384, 1086

\bibitem[Joarder \& Satya Narayanan(2000)]{joardersatya} Joarder, P. S., \& Satya Narayanan, A. 2000, \aap, 359, 1211

\bibitem[Joarder \& Nakariakov(2006)]{joardernaka} Joarder, P. S., \& Nakariakov, V. M. 2006, Geophys. Astrophys. Fuid Dynamics, 100, 59

\bibitem[Kolesnikov et al.(2004)]{kolesnikov} Kolesnikov, F., B\"unte, M., Schmitt, D., \& Sch\"ussler, M. 2004, \aap, 420, 737

\bibitem[Lin et al.(2009)]{lin} Lin, Y., Soler, R., Engvold, O., Ballester, J. L., Langangen, \O., Oliver, R., \& Rouppe van der Voort, L. H. M. 2009, \apj, 704, 870

\bibitem[Nakariakov et al.(1999)]{naka1} Nakariakov V. M., Ofman, L., Deluca, E. E., Roberts, B., Davila, J. M. 1999, Science, 285, 862

\bibitem[Nakariakov \& Ofman(2001)]{naka2} Nakariakov V. M., \& Ofman, L. 2001, \aap, 372, L53

\bibitem[Ofman(2009)]{ofman09} Ofman, L. 2009, \ssr, in press

\bibitem[Ogilvie \& Pringle(1996)]{ogilvie} Ogilvie, G. I., \& Pringle, J. E. 1996, \mnras, 279, 152 

 \bibitem[Okamoto et al.(2007)]{okamoto} Okamoto, T. J, et al. 2007, Science, 318, 1557 


\bibitem[Rae(1983)]{rae} Rae, I. C. 1983, \aap, 126, 209

\bibitem[Rae \& Roberts(1983)]{raeroberts} Rae, I. C., \& Roberts, B. 1983, \solphys, 84, 99

\bibitem[Ruderman(2007)]{ruderman} Ruderman, M. S. 2007, \solphys, 246, 119

\bibitem[Ruderman \& Roberts(2002)]{rudermanroberts} Ruderman, M., \& Roberts, B. 2002, \apj, 577, 475

\bibitem[Ruderman \& Roberts(2006)]{rudermanroberts06} Ruderman, M., \& Roberts, B. 2006, J. Plasma Physics, 72, 285

\bibitem[Ruderman \& Erd\'elyi(2009)]{rudermanrobertus} Ruderman, M., \& Erd\'elyi, R. 2009, \ssr, in press



    \bibitem[Soler et al.(2008)]{solerflow} Soler, R., Oliver, R., \& Ballester, J. L. 2008,  \apj, 684, 725 
  \bibitem[Soler et al.(2009)]{solermulti} Soler, R., Oliver, R., \& Ballester, J. L. 2009,  \apj, 693, 1601 

\bibitem[Erd\'elyi \& Taroyan(2003)]{taroyan} Erd\'elyi, R., \& Taroyan, Y. 2003, Journal Geophys. Res., 108, 1043

\bibitem[Terra-Homem et al.(2003)]{terra} Terra-Homem, M., Erd\'elyi, R., \& Ballai, I. 2003, \solphys, 217, 199

\bibitem[Terradas et al.(2007)]{terradasenergy} Terradas, J., Andries, J., \& Goossens, M. 2007, \apj, 469, 1135

 \bibitem[Terradas et al.(2008a)]{hinode} Terradas, J., Arregui, I., Oliver, R., \& Ballester, J. L. 2008a, \apj, 678, L153

\bibitem[Terradas et al.(2008b)]{terradas} Terradas, J., Andries, J., Goossens, M., Arregui, I., Oliver, R., \& Ballester, J. L. 2008b, \apj, 687, L115

\bibitem[Terradas et al.(2009)]{terradasflow} Terradas, J., Goossens, M., \& Ballai, I. 2009, \aap, submitted

\bibitem[Terradas(2009)]{terradasreview} Terradas, J. 2009, \ssr, in press

\bibitem[Wilson(1981)]{wilson} Wilson, P. R. 1981, \apj, 251, 756







  

%
 \end{thebibliography}
\end{document}